\documentclass[conference]{IEEEtran}
\IEEEoverridecommandlockouts

\usepackage{cite}
\usepackage{amsmath,amssymb,amsfonts}
\usepackage{algorithmic}
\usepackage{graphicx}
\usepackage{textcomp}
\usepackage{xcolor}
\usepackage{hyperref}
\usepackage{multirow}
\usepackage{subcaption}

\def\BibTeX{{\rm B\kern-.05em{\sc i\kern-.025em b}\kern-.08em
    T\kern-.1667em\lower.7ex\hbox{E}\kern-.125emX}}

\usepackage[export]{adjustbox}

\makeatletter
\newcommand{\newlineauthors}{%
  \end{@IEEEauthorhalign}\hfill\mbox{}\par
  \mbox{}\hfill\begin{@IEEEauthorhalign}
}
\makeatother

\title{A Mobile Application Front-End for Presenting Explainable AI Results in Diabetes Risk Estimation}

\author{
\IEEEauthorblockN{Bernardus Willson}
\IEEEauthorblockA{School of Electrical Engineering and \\ Informatics \\
Bandung Institute of Technology \\
Bandung, Indonesia \\
13521021@std.stei.itb.ac.id}
\and
\IEEEauthorblockN{Henry Anand Septian Radityo}
\IEEEauthorblockA{School of Electrical Engineering and \\ Informatics \\
Bandung Institute of Technology \\
Bandung, Indonesia \\
13521004@std.stei.itb.ac.id}
\and
\IEEEauthorblockN{Raynard Tanadi}
\IEEEauthorblockA{School of Electrical Engineering and \\ Informatics \\
Bandung Institute of Technology \\
Bandung, Indonesia \\
13521143@std.stei.itb.ac.id}
\newlineauthors
\IEEEauthorblockN{Latifa Dwiyanti}
\IEEEauthorblockA{School of Electrical Engineering and Informatics \\
Bandung Institute of Technology \\
Bandung, Indonesia \\
latifa@informatika.org}
\and
\IEEEauthorblockN{Saiful Akbar}
\IEEEauthorblockA{School of Electrical Engineering and Informatics \\
Bandung Institute of Technology \\
Bandung, Indonesia \\
saiful@informatika.org}
}

\begin{document}

\maketitle

\IEEEoverridecommandlockouts
\IEEEpubid{\makebox[\columnwidth][l]{979-8-3315-7578-6/25/\$31.00~\copyright~2025 IEEE} \hspace{\columnsep}\makebox[\columnwidth]{ }}
\IEEEpubidadjcol

\begin{abstract}
Diabetes is a significant and continuously rising health challenge in Indonesia. Although many artificial intelligence (AI)-based health applications have been developed for early detection, most function as "black boxes," lacking transparency in their predictions. Explainable AI (XAI) methods offer a solution, yet their technical outputs are often incomprehensible to non-expert users. This research aims to develop a mobile application front-end that presents XAI-driven diabetes risk analysis in an intuitive, understandable format. Development followed the waterfall methodology, comprising requirements analysis, interface design, implementation, and evaluation. Based on user preference surveys, the application adopts two primary visualization types—bar charts and pie charts—to convey the contribution of each risk factor. These are complemented by personalized textual narratives generated via integration with GPT-4o. The application was developed natively for Android using Kotlin and Jetpack Compose.
The resulting prototype interprets SHAP (SHapley Additive exPlanations)—a key XAI approach—data into accessible graphical visualizations and narratives. Evaluation through user comprehension testing (Likert scale and interviews) and technical functionality testing confirmed the research objectives were met. The combination of visualization and textual narrative effectively enhanced user understanding (average score 4.31/5) and empowered preventive action, supported by a 100\% technical testing success rate.
\end{abstract}


\section{Introduction}
Diabetes is a major public health issue in Indonesia, ranking fifth globally with 19.5 million sufferers in 2021, a figure projected to reach 28.6 million by 2045 \cite{b1}. This trend poses a significant threat to individual health and imposes a substantial burden on the national healthcare system and economy. Early detection is therefore critical for implementing preventive measures before the disease progresses.

Artificial Intelligence (AI) has shown great potential in enhancing the accuracy and efficiency of disease detection, including diabetes \cite{b2}. Explainable Artificial Intelligence (XAI) is a promising subfield that clarifies the reasoning behind predictions, enhancing transparency \cite{b3}. This is crucial in healthcare, where decisions can significantly impact a person's quality of life. However, a key challenge is that XAI outputs are often too technical for non-expert users to interpret \cite{b4}. This communication gap can lead to confusion and mistrust, undermining the goal of empowering users with actionable health insights. Therefore, a critical need exists for an interface that translates these technical explanations, such as SHAP values, into an intuitive format for end-users.

To address this challenge, this paper details the development of a front-end mobile application designed to bridge the gap between complex XAI outputs and user understanding. This work makes two primary contributions. The first is the design of a mobile user interface that effectively visualizes and communicates XAI-driven diabetes risk explanations to a general Indonesian audience, a crucial step for improving user trust and health literacy in a region significantly impacted by the disease. Second, we developed a robust presentation component to process and display complex XAI results, ensuring a seamless and reliable user experience.

\section{Related Work}
The challenge of "black box" models in health AI has spurred the adoption of Explainable AI (XAI) methods, such as SHAP (SHapley Additive exPlanations) and LIME (Local Interpretable Model-agnostic Explanations), to enhance transparency \cite{b5}. However, research indicates that this solution introduces its own challenge: the outputs from these methods are often too technical for the intended end-users, namely patients and the general public \cite{b6}. This "comprehension gap" is a recognized barrier, as presenting complex data without proper translation can lead to user misinterpretation—a significant risk in critical healthcare applications where clear understanding is paramount for informed decision-making \cite{b7}.

To bridge this gap, research has focused on translating XAI outputs into more user-friendly formats, mainly through data visualization. Although libraries like SHAP and LIME offer built-in visualizations such as waterfall, force, and bar plots \cite{b8}, their technical nature often imposes a high cognitive load on non-expert users. In contrast, familiar formats like bar and pie charts are easier to interpret because they reduce cognitive effort \cite{b9}. Large Language Models (LLMs) further enhance accessibility by converting technical XAI outputs into natural, human-readable narratives \cite{b10}. Prior studies show that combining visualizations with LLM-generated explanations yields the highest user understanding, outperforming both raw AI predictions and standard XAI plots \cite{b11}.

Existing AI-based diabetes management applications, such as \textit{DiabTrend}, often operate as "black boxes," providing predictions without explaining the underlying rationale. This lack of transparency, frequently coupled with poor usability from cluttered interfaces, fails to effectively communicate risk factors to users. As a result, users are limited in their ability to take informed preventive action. This highlights a clear need for a user-centered application that delivers intuitive, accessible, and actionable risk explanations, which is the primary motivation for this research.

\section{Methodology}
\begin{figure}[h!]
     \centering
     \includegraphics[width=0.9\columnwidth]{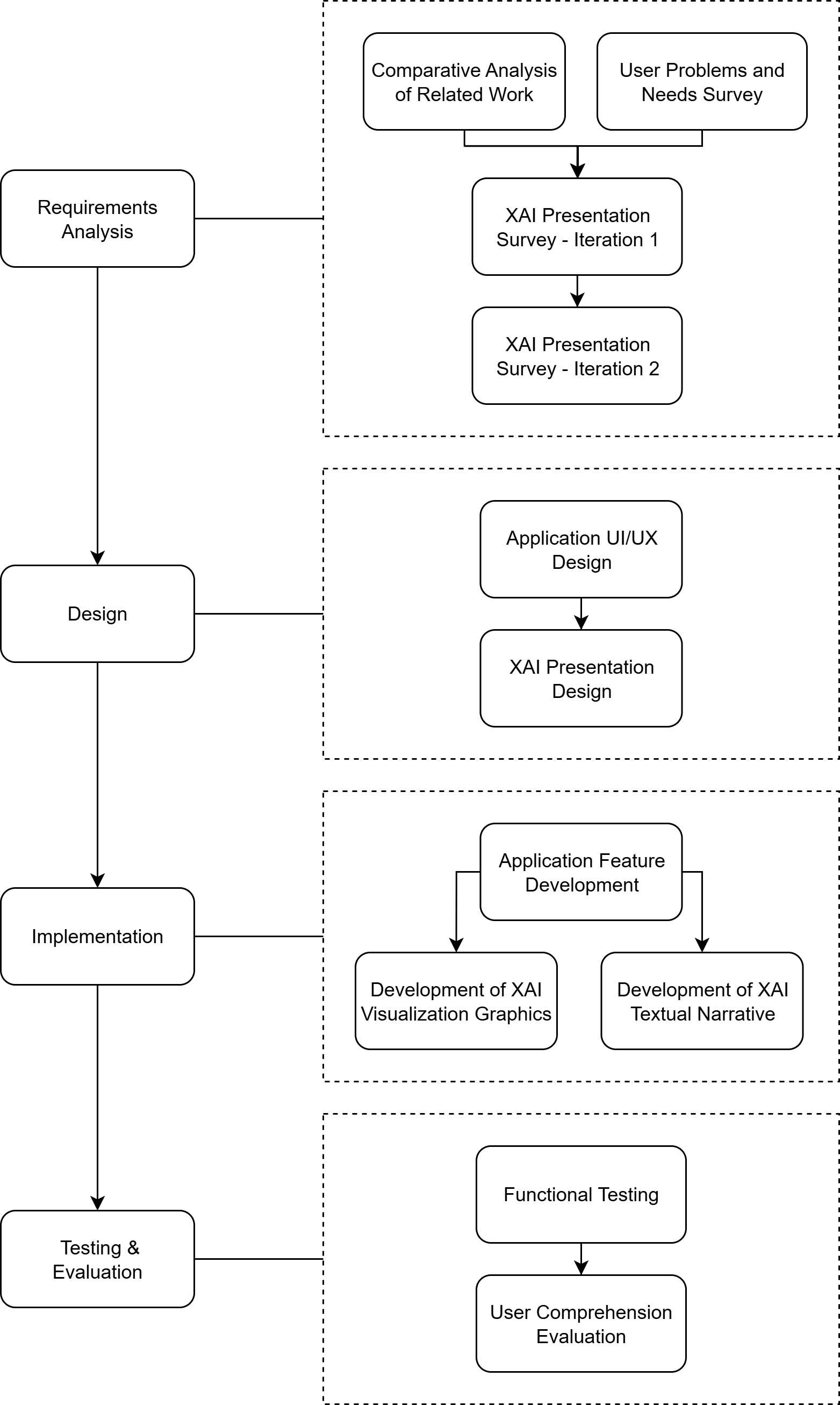}
     \caption{Flowchart of the Research Methodology.}
     \label{fig:methodology_flowchart}
 \end{figure}

The project followed the waterfall model, a sequential software development life cycle (SDLC) where distinct phases are completed in a strict linear sequence \cite{b12}. The scope of this research covers the phases from requirements analysis to testing and evaluation; deployment and maintenance are considered outside the project's scope. The development process is outlined in Figure \ref{fig:methodology_flowchart}.

\subsection{Requirements Analysis}
The initial phase involved a public survey to analyze the needs and preferences of potential users. Alongside the survey, a comparative analysis of related work was conducted to identify existing solutions, identify unmet needs, and establish a feature baseline. Data was collected on key health topics, preferred presentation formats, and desired application features. This phase produced a user requirements specification document to guide the subsequent design.

\subsection{Design}
Based on the requirements analysis, this phase focused on designing the application's concept and architecture. This included UI/UX design, creating an initial prototype focused on usability and aesthetics, and designing the presentation format for XAI results to ensure they were easily understood by users.

\subsection{Implementation}
Following the design phase, the application was built. The design was implemented in code using the designated technologies: Kotlin and Android Studio. In this stage, all features were developed and integrated with back-end services and the machine learning model to ensure accurate risk analysis presentation.

\subsection{Testing and Evaluation}
The implemented application underwent a comprehensive evaluation, beginning with functionality testing to verify all features and eliminate critical bugs. Subsequently, user testing was conducted via questionnaires and interviews to gather feedback on user experience and the comprehensibility of XAI results, which informed future development.

\section{Requirements Analysis and Key Findings}
The requirements analysis phase was conducted through a multi-stage user survey process designed to identify user challenges and determine the most effective methods for presenting XAI results. Three distinct surveys were administered, each with a specific objective, number of respondents, and scope. A summary of this process is presented in Table~\ref{tab:survey_summary}.

\begin{table}[h!]
\caption{Summary of User Surveys Conducted}
\label{tab:survey_summary}
\centering
\begin{tabular}{|c|c|p{4.5cm}|}
\hline
\textbf{Survey} & \textbf{Respondents} & \textbf{Objective} \\
\hline
1 & 333 & To identify primary user challenges in understanding and trusting AI-driven health information. \\
\hline
2 & 183 & To compare the comprehension of default XAI library plots against custom-simplified visualizations. \\
\hline
3 & 114 & To refine the choice of visualization by comparing the three most effective simplified chart formats. \\
\hline
\end{tabular}
\end{table}

The initial survey (N=333) identified three primary user challenges: uncertainty regarding the accuracy of health information, difficulty obtaining relevant health recommendations, and difficulty understanding complex data presentations.

\begin{figure}[htbp]
    \centering
    \begin{subfigure}[b]{0.48\columnwidth}
        \includegraphics[width=\textwidth]{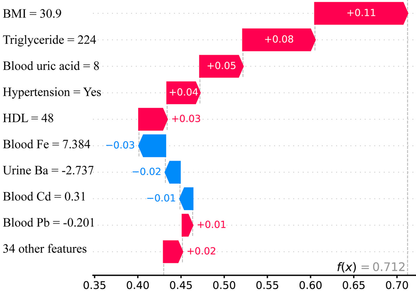}
        \caption{Example of a technical default plot (SHAP Waterfall).}
        \label{fig:default_xai}
    \end{subfigure}
    \hfill
    \begin{subfigure}[b]{0.48\columnwidth}
        \includegraphics[width=\textwidth]{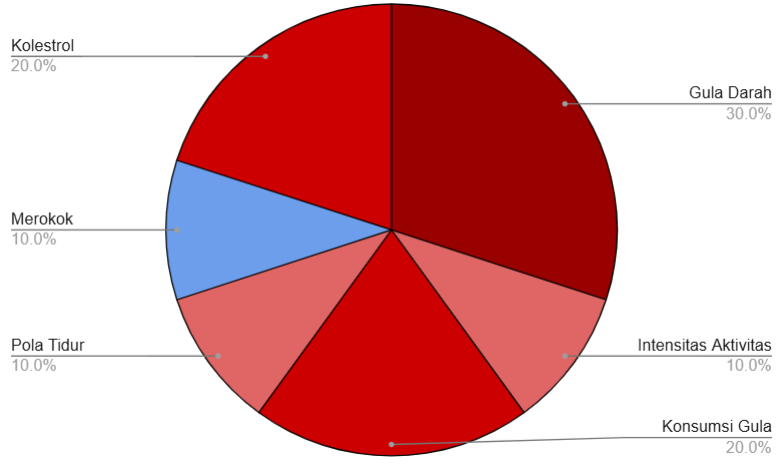}
        \caption{Example of a simplified, user-preferred format (Pie Chart).}
        \label{fig:custom_xai}
    \end{subfigure}
    \caption{Comparison between a standard technical XAI output and the simplified, user-preferred formats developed in this research.}
    \label{fig:xai_comparison}
\end{figure}

To directly address the challenge of data presentation, a second survey (N=183) was conducted. This study compared user comprehension of default plots from common XAI libraries (a LIME bar plot, a SHAP bar plot, a SHAP waterfall plot, and a SHAP force plot) with custom-designed, simplified graphics (a pie chart, a bar chart, and a radar chart), as illustrated in Fig.~\ref{fig:xai_comparison}. The findings revealed a clear user preference for the simplified visualizations. The pie chart and bar chart achieved the highest comprehension scores (8.69 and 8.54 out of 10, respectively), significantly outperforming the more technical XAI outputs. A key finding was that 85.2\% of respondents considered narrative text essential for providing context.

Building on these results, a third survey (N=114) sought to determine the most preferred format among the top-performing simplified charts. This survey compared three specific formats: a pie chart, a standard bar chart (with all bars extending to the right), and a diverging bar chart (with bars extending left or right to show positive/negative influence). Although preferences were closely divided, the standard bar chart received the most votes (39.5\%), with the diverging bar chart emerging as a strong second choice. This final survey reinforced the critical need for supporting textual explanations to define terms and clarify the magnitude and direction of each risk factor's influence. Collectively, these findings formed the core design principles for the application's user interface.

\section{System Design}

\subsection{System Functionality}
The application's functionalities are designed to provide users with accurate risk estimations accompanied by understandable explanations. The key features are summarized in Table \ref{tab:system_functionalities}.
\begin{table}[h!]
\caption{Summary of System Functionalities}
\label{tab:system_functionalities}
\centering
\begin{tabular}{|p{2cm}|p{2.5cm}|p{3.5cm}|}
\hline
\textbf{Category} & \textbf{Feature} & \textbf{Functional Description} \\
\hline
\multirow{4}{2cm}{Registration \& Access}
    & Onboarding & Introduces main features on first launch. \\
\cline{2-3}
    & User Registration & New user sign-up with email and OTP verification. \\
\cline{2-3}
    & User Login & Authenticates and grants access to existing users. \\
\cline{2-3}
    & Password Recovery & Enables password reset via an email OTP. \\
\hline
\multirow{4}{2cm}{Estimation \& Education}
    & Initial Survey & Collects initial health data for the first risk assessment. \\
\cline{2-3}
    & AI Risk Estimation & Calculates and displays diabetes risk percentage and level. \\
\cline{2-3}
    & Risk Factor Explanation (XAI) & Visually explains key factors influencing the risk score. \\
\cline{2-3}
    & Lifestyle Change Simulation & Simulates how lifestyle changes affect the risk percentage. \\
\hline
\multirow{3}{2cm}{Monitoring \& Data}
    & Daily/Non-Daily Data Logging & Allows users to log health and lifestyle data regularly. \\
\cline{2-3}
    & Risk History Graph & Shows a 30-day trend graph of the user's risk score. \\
\cline{2-3}
    & Profile Update & Allows users to edit their personal profile information. \\
\hline
\multirow{3}{2cm}{Supporting Features}
    & Information \& Help & Offers app info, prevention tips, and an FAQ section. \\
\cline{2-3}
    & Reminder Notification & Sends daily reminders to log health data. \\
\cline{2-3}
    & Connection Detection & Alerts user when there is no internet connection. \\
\hline
\end{tabular}
\end{table}

\subsection{XAI Result Presentation}

Based on survey findings, our approach to presenting the XAI results is founded on two core design principles: offering clear, alternative visualizations and supplementing them with concise, narrative explanations.

First, recognizing that no single visualization is optimal for all user segments, the application provides two distinct options to accommodate diverse preferences: a bar chart and a pie chart. These visualizations were custom-designed for clarity and simplicity, intentionally avoiding the complexity of default XAI library plots. The magnitude of each risk factor's contribution is intuitively represented by the length of the bar or the angle of the pie slice. To ensure clarity, a custom legend accompanies each chart, displaying numerical percentages and abbreviations for each risk factor. To further enhance immediate comprehension, a color-coded system is employed: red is used to highlight factors that increase risk, while green signifies factors that decrease risk.

Second, acknowledging that visualizations alone are insufficient for a comprehensive understanding, each graphic is paired with a supporting narrative text. This narrative is designed to explain three key points:
\begin{itemize}
    \item Whether a given factor increases/decreases the user's risk.
    \item The magnitude of that factor's influence.
    \item A brief definition of the factor itself.
\end{itemize}

To achieve this, the application leverages a Large Language Model (LLM) to translate technical data into simple, easy-to-understand language, with each explanation crafted into two to three concise sentences. This approach aligns with research indicating that LLMs can effectively translate technical SHAP values into accessible natural language \cite{b20}.

\begin{figure}[htbp]
\centerline{\includegraphics[width=\columnwidth]{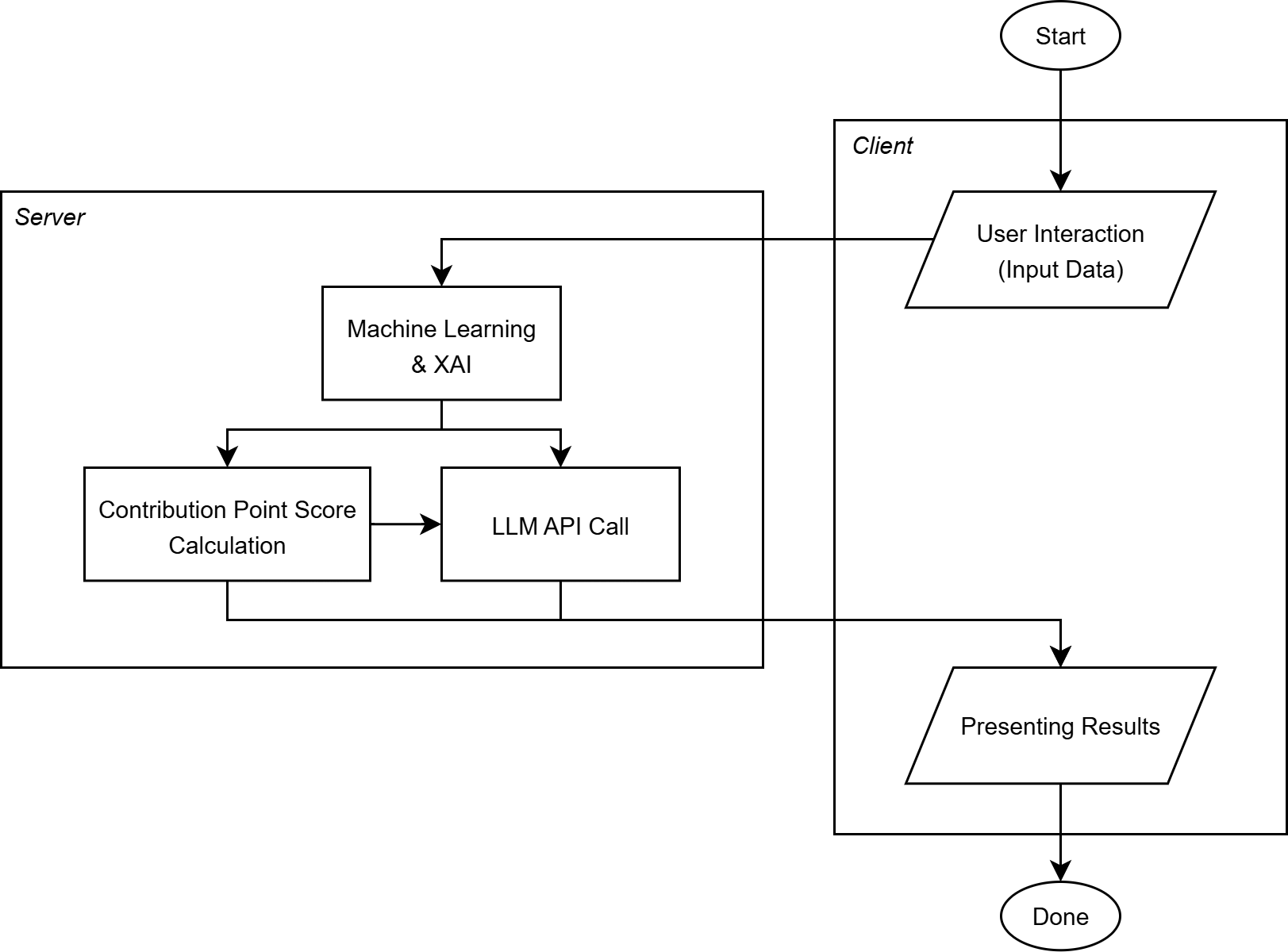}}
\caption{System Workflow for Processing and Presenting XAI Results.}
\label{fig:flow}
\end{figure}

The overall data flow for presenting these XAI results, as illustrated in Fig.~\ref{fig:flow}, begins on the client-side. The user interacts with the application interface to input their relevant health data. This data is then sent to the server, where machine learning and XAI models generate a risk estimation and identify the most influential factors. Subsequently, the server performs two parallel processes: it calculates the percentage contribution of each factor for visualization and calls the LLM API to generate the personalized textual narrative. Once the quantitative data and the narrative explanations are prepared, the combined results are sent back to the client. Finally, the application presents this integrated information to the user through their chosen visualization, displayed alongside the explanatory text.

\section{Implementation}

\subsection{Development Environment and Technologies}
The application was developed natively for the Android platform using Kotlin, chosen for its optimal performance \cite{b13}. The user interface was built with Jetpack Compose, a modern UI toolkit that simplifies front-end development. For data visualization, the MPAndroidChart library was used \cite{b14}. The back-end system utilized the OpenAI API with the GPT-4o model to generate narrative explanations, a decision made to align with user preferences for custom textual and visual outputs \cite{b15}.

\subsection{Implementation of XAI Graphic Visualization}
\begin{figure}[htbp]
\centerline{\includegraphics[width=0.5\textwidth]{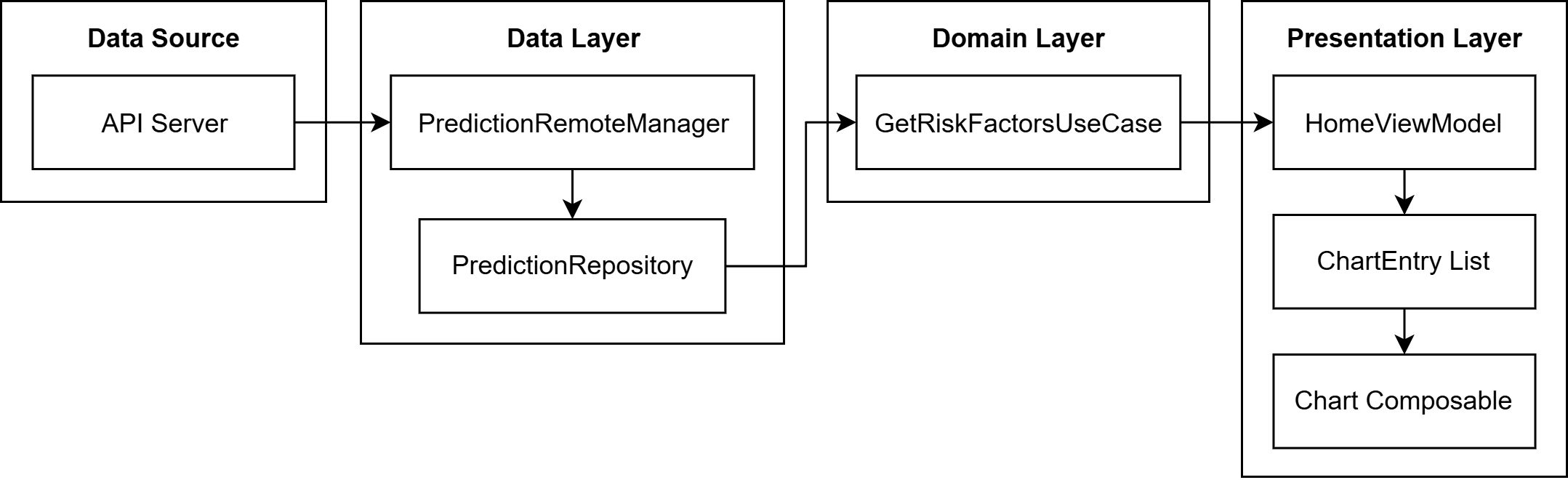}}
\caption{Data Flow for Graphic Visualization.}
\label{fig:data_flow}
\end{figure}

The implementation of the bar and pie charts was accomplished using the MPAndroidChart library, integrated into Jetpack Compose via the AndroidView component. This approach is situated within an MVVM (Model-View-ViewModel) and Clean Architecture framework, ensuring a clear separation of concerns as data flows from the remote data source to the presentation layer, as illustrated in Figure \ref{fig:data_flow}.

\begin{figure}[htbp]
\centerline{\includegraphics[width=0.4\textwidth, frame]{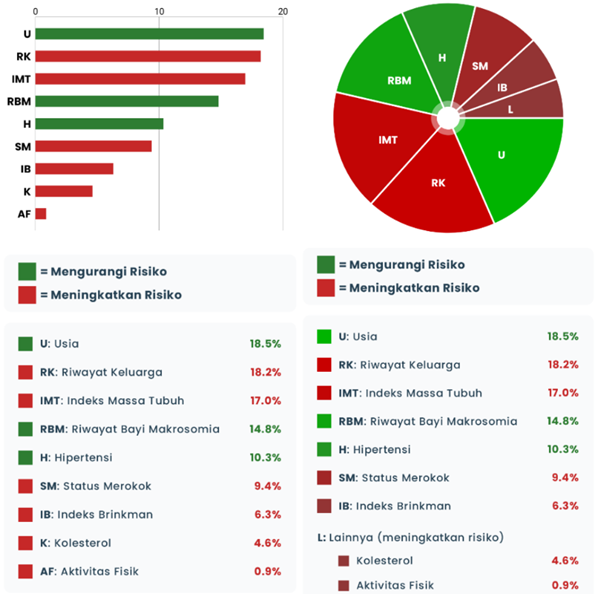}}
\caption{Example of Bar Chart and Pie Chart visualizations.}
\label{fig:charts}
\end{figure}

To make the data more intuitive for the end-user, the raw SHAP values from the XGBoost model were converted into percentages using the formula shown in Equation (\ref{eq:shap_percentage}). Figure \ref{fig:charts} provides an example of this personalized output, illustrating how the model can identify specific factors like 'Family History' as a primary risk driver for a user, while simultaneously showing how other factors, such as 'Age', may be reducing their risk.

\begin{equation} \label{eq:shap_percentage}
P_i = \frac{|S_i|}{\sum_{j=1}^{n} |S_j|} \times 100\%
\end{equation}
where $P_i$ is the percentage contribution of feature $i$, and $S_i$ is its SHAP value.

\subsection{Implementation of Textual Narrative with LLM}
To generate factually grounded narratives and mitigate the risk of model hallucination, a systematic prompt engineering process was employed using the GPT-4o model. This involved a comprehensive 'system prompt' that defined the model’s behavior and a 'user prompt' containing the specific patient data for each request. Key strategies from established prompt engineering practices were integrated into the system prompt, including defining a persona, specifying the task, providing context, and using 'few-shot' examples \cite{b16, b17}. The prompt established a 'Medical AI Explainer' persona, defined the core task, and enforced a strict JSON output structure. To ground the model's responses, it was provided with a knowledge base containing feature definitions and global feature importance statistics. The prompt itself was structured using Markdown, leveraging tables to improve model comprehension, a technique recommended by recent studies \cite{b18, b19}.

\begin{figure}[htbp]
\centerline{\includegraphics[width=0.7\columnwidth]{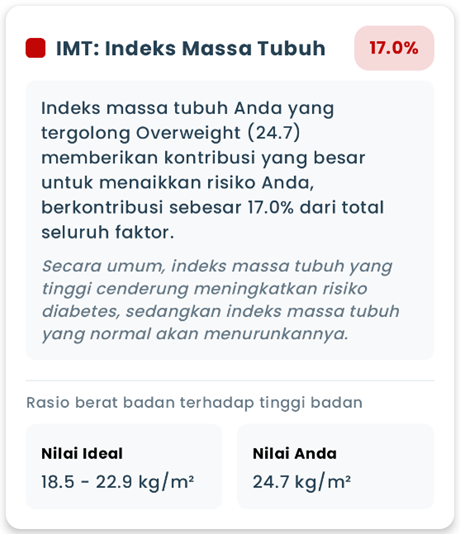}}
\caption{Example of a narrative explanation card for BMI.}
\label{fig:narrative-card}
\end{figure}

The resulting JSON was then parsed and displayed in the app within a detailed narrative card for each risk factor. An example of this component is shown for the Body Mass Index (BMI) factor in Figure~\ref{fig:narrative-card}. The title indicates BMI's contribution to the total risk influence, which for this user is 17.0\%. The primary explanation, a two-sentence narrative, is dynamically generated by the LLM: the first sentence describes the personal impact (an overweight BMI of 24.7), while the second provides general context on how BMI typically relates to diabetes risk. To further ground the user's understanding, the card concludes with a direct comparison between the user's value (`Nilai Anda`), pulled from user data, and the standard ideal range (`Nilai Ideal`), which is based on static definitions.

\section{Testing and Evaluation}
The application prototype was subjected to a two-pronged evaluation strategy: a user comprehension study to assess the clarity and usefulness of the XAI presentation, and a technical evaluation to validate the application's functional reliability.

\subsection{Functional Evaluation}
To validate the application's technical robustness, we conducted 111 automated end-to-end (E2E) test scenarios using the Espresso framework, with the results summarized in Table \ref{table:e2e_results}. These tests covered all major user workflows, including registration, data input, risk simulation, and viewing results, under both ideal ("happy path") and negative conditions. The testing achieved a 100\% pass rate, exceeding the 90\% target and confirming that the application is technically stable, reliable, and functions according to specifications.

\begin{table}[h!]
\caption{Summary of End-to-End (E2E) Functional Test Results}
\label{table:e2e_results}
\centering
\begin{tabular}{|l|c|c|c|}
\hline
\textbf{Use Case Group} & \textbf{Scenarios} & \textbf{Pass} & \textbf{Fail} \\
\hline
Onboarding & 3 & 3 & 0 \\
Registration & 13 & 13 & 0 \\
Password Change & 7 & 7 & 0 \\
Login & 8 & 8 & 0 \\
Initial Survey & 18 & 18 & 0 \\
Add Daily Data & 6 & 6 & 0 \\
Update Non-Daily Data & 18 & 18 & 0 \\
View Risk Estimation & 4 & 4 & 0 \\
View Risk Factor Explanation & 4 & 4 & 0 \\
Simulate Risk Scenario & 13 & 13 & 0 \\
View Today's Data & 1 & 1 & 0 \\
View History & 5 & 5 & 0 \\
View Guide \& Tips & 1 & 1 & 0 \\
Update User Info & 4 & 4 & 0 \\
Connectivity Alerts & 5 & 5 & 0 \\
Logout & 1 & 1 & 0 \\
\hline
\textbf{Total} & \textbf{111} & \textbf{111} & \textbf{0} \\
\hline
\end{tabular}
\end{table}

\subsection{User Comprehension Evaluation}
We evaluated the prototype with 12 participants from diverse educational backgrounds and ages (20--64 years). The evaluation employed a mixed-methods approach, combining a 5-point Likert scale questionnaire with semi-structured interviews to gather both quantitative and qualitative data.

As shown in Table~\ref{table:likert_results}, the overall mean score for user comprehension was 4.31 out of 5 ($SD=0.88$), significantly surpassing our target of 4.0 and indicating a high level of user satisfaction and understanding. The highest-rated aspect was the combination of graphics and text (mean score of 4.67), confirming that this synergistic approach was the most effective design choice.

\begin{table}[h!]
\caption{Likert Scale Results for User Comprehension (N=12)}
\label{table:likert_results}
\centering
\begin{tabular}{|l|c|c|}
\hline
\textbf{Statement Category} & \textbf{Mean Score} & \textbf{Std. Dev.} \\
\hline
Clarity of Bar Chart & 3.75 & 1.09 \\
Clarity of Pie Chart & 3.75 & 1.01 \\
Ease of Comparing Factors (Graphics) & 4.42 & 0.86 \\
Clarity of Textual Explanation & 4.50 & 0.65 \\
Simplicity of Language & 4.58 & 0.49 \\
\textbf{Helpfulness of Graph + Text Combo} & \textbf{4.67} & \textbf{0.62} \\
Understanding of Personal Risk Score & 4.42 & 0.95 \\
Empowerment for Preventive Action & 4.42 & 0.64 \\
\hline
\textbf{Overall Average} & \textbf{4.31} & \textbf{0.88} \\
\hline
\end{tabular}
\end{table}

A deeper analysis of the results—particularly the identical low scores and high standard deviations for the individual charts—revealed nuanced user preferences linked to demographic backgrounds.

\begin{itemize}
    \item \textbf{Influence of Educational and Professional Background:} A distinct pattern emerged where participants with backgrounds in fields requiring structured data analysis, such as Accounting and Business, consistently preferred the bar chart for its functional clarity and precise visual comparison. Conversely, the pie chart was favored by participants who prioritized aesthetics, including those from technical fields like Informatics and Architecture, as well as those with lower educational attainment.
    \item \textbf{Impact of Data Literacy:} Participants with lower educational attainment (e.g., elementary or middle school graduates) initially found both graphs challenging to understand on their own. However, they could still extract the primary message by relying on fundamental visual cues, such as the color red indicating danger and green indicating good. 
    For this group, the narrative text was essential for converting visual data into concrete understanding.
    \item \textbf{Expert-Driven Feedback:} A participant with a Design Communication Visuals background offered expert insights, noting that combining positive (risk-decreasing) and negative (risk-increasing) factors in one graphic reduces clarity. They recommended using a diverging bar chart, which aligns with findings from our second user survey. The participant also suggested distinguishing controllable (e.g., lifestyle) from uncontrollable (e.g., genetic) factors to provide more actionable results.
\end{itemize}
Ultimately, the combination of graphics and text proved to be a universally effective and inclusive strategy. It allowed all participants to synthesize the information, understand their personal risk profile, and formulate concrete preventive plans, such as improving diet and increasing physical activity.

\section{Conclusion}
This paper details the development of a mobile front-end that translates complex, technical Explainable AI (XAI) outputs into actionable health insights for lay users. Our approach, which combines simple, user-preferred visualizations (bar and pie charts) with personalized, LLM-generated narratives, proved highly effective in making diabetes risk predictions both accessible and understandable.

The success of this approach is confirmed by our two-pronged evaluation. Quantitatively, the application achieved a high user comprehension score of 4.31 out of 5 and a 100\% pass rate in functional end-to-end testing, validating both its usability and technical robustness. Qualitatively, our analysis revealed that the core strength of the design lies in the synergistic combination of graphics and text. This pairing was universally praised by participants, proving to be an inclusive method that successfully bridged comprehension gaps across diverse educational backgrounds and levels of data literacy.

Our user evaluation also illuminated promising directions for future work. First, in response to expert feedback on visual clarity, future iterations could explore alternative visualizations, such as a diverging bar chart, to more intuitively separate risk-increasing from risk-decreasing factors. Second, a strategic recommendation from our study was to restructure outputs by separating controllable (e.g., lifestyle) from uncontrollable factors (e.g., genetics), shifting the application from diagnosis to a proactive health guide that emphasizes actionable steps.


\end{document}